\documentclass[12pt]{article}

\usepackage[textheight=22.5truecm,textwidth=16.8truecm,top=2.95truecm,left=2.15truecm]{geometry}
\usepackage{
amsmath,
amsthm,
amssymb,
ascmac,
bm,
tikz,
braket,
mathrsfs,
graphicx,
color,
hyperref,
cite,
titlesec
}

\hypersetup{colorlinks,bookmarksopen,bookmarksnumbered,citecolor=black,linkcolor=black,linktocpage,pdfstartview=FitH,urlcolor=black}

\numberwithin{equation}{section}

{\makeatletter
\long\def\@makefntext#1{\parindent 1em\noindent 
\@hangfrom{\hbox to 1.8em{\hss$^{\@thefnmark}$}}#1}
\makeatother}

\def\fnum@figure{\textbf{\figurename\nobreakspace\thefigure}}
\def\fnum@table{\textbf{\tablename\nobreakspace\thetable}}
\long\def\@makecaption#1#2{%
  \vskip\abovecaptionskip
  \sbox\@tempboxa{\small #1. #2}%
  \ifdim \wd\@tempboxa >\hsize
    \small #1. #2\par
  \else
    \global \@minipagefalse
    \hb@xt@\hsize{\hfil\box\@tempboxa\hfil}%
  \fi
  \vskip\belowcaptionskip}

\def\l{\left}
\def\r{\right}

\def\d{\mathrm{d}}

\title{\hfill\parbox{3cm}{\normalsize KUNS-2909}\\[12pt]
Light-cone cuts and hole-ography:\\ explicit reconstruction of bulk metrics
}
\author{
Daichi Takeda\footnote{takedai@gauge.scphys.kyoto-u.ac.jp}\\[12pt]
 \textit{Department of Physics, Kyoto University, Kyoto 606-8502, Japan}
}
\date{}

\begin{document}
\maketitle
\begin{abstract}
	\noindent
	In this paper, the two reconstruction methods, light-cone cuts method and hole-ography, are
	combined to provide complete bulk metrics of locally AdS$_3$ static spacetimes.
	As examples, our method is applied to the geometries of pure AdS$_3$, AdS$_3$ soliton, 
	and BTZ black hole,
	and we see them successfully reconstructed.
	The light-cone cuts method is known to have difficulty in obtaining conformal factors,
	while the hole-ography in describing temporal components.
	Combining the two methods, we overcome the disadvantages and give complete metrics 
	for a class of holographic theories such that entanglement wedge and causal wedge coincide.
	Light-cone cuts are identified by entanglement entropy in our method.
	We expect our study to lead to the discovery of a universal relation between the two methods,
	by which the combination would be applied to more generic cases.
\end{abstract}
\newpage
\tableofcontents

\section{Introduction}\label{sec: introduction}
In gauge/gravity duality \cite{Maldacena:1997re,Witten:1998qj,Gubser:1998bc},
the emergence of spacetimes from quantum field theories is a subject to be addressed.
It implies that quantum effects on gravity are encoded in dynamics of quantities in quantum field theories,
indicating the possibility of describing quantum gravity.
In order to unravel the mysterious mechanism, we should first start with investigating the mechanism 
at the classical level.

There are various attempts to construct spacetime metrics in AdS/CFT correspondence, and great progress has
been made so far.
In such studies,  a reconstruction method using light-cone cuts,\footnote{
A light-cone cut is the intersection between the conformal boundary and the light-cone of a certain bulk point.
A more precise definition is introduced later.
}
was proposed by Engelhardt and Horowitz \cite{Engelhardt:2016wgb}.
As a boundary observable, a specific divergence of correlators called ``bulk-point singularity"
\cite{Maldacena:2015iua} is used to find light-cone cuts.
It is notable that the method can be applied to variety of holographic theories,
for it does not assume symmetries, matter contents,
and the bulk equation of motion, and is supported by mathematical proofs.
In addition, the reconstruction procedure after obtaining light-cone cuts
is relatively practical, which is demonstrated in \cite{Burda:2018rpb}.
The extension to spacetimes with a compact space is also examined in \cite{Hernandez-Cuenca:2020ppu}.

However, the light-cone cuts method alone cannot determine local conformal factors,
for the set of light-cone cuts is identical between geometries having the same causal structure.
The determination of conformal factors with additional ingredients is discussed 
in \cite{Engelhardt:2016crc} for two cases:
one is when some information about the matter is known, and the other is when the bulk is a small perturbation
of pure AdS.
In the former, one component of Einstein equation is assumed, and in the latter, 
boundary entanglement entropy is used.
Other than that, constraints on conformal factors were discussed recently in \cite{Folkestad:2021kyz},
by using curvature conditions.
As another concern, the bulk-point singularity is not suitable for finding light-cone cuts corresponding to points 
near the horizon of an eternal black hole \cite{Engelhardt:2016wgb,Engelhardt:2016crc}, and
finding light-cone cuts by the singularity is a tough computation.
As is also mentioned in their paper, we would like to expect that there would exist better ways to find light-cone cuts.
In this paper, we give some progress on the two issues about conformal factors and finding 
light-cone cuts.

Then, what boundary quantity would help us solve the problem about conformal factors?
In AdS/CFT correspondence, holographic entanglement entropy formula \cite{Ryu:2006bv,Ryu:2006ef,Hubeny:2007xt}
is known to be a useful tool in reconstructing bulk.
The formula has made it turn out that boundary entanglement entropy is deeply related to the bulk geometry,
and motivated by the fact, various proposals of bulk reconstruction have been made so far.
Thus, entanglement entropy has a potential to reconstruct conformal factors.

In such proposals, we focus on a reconstruction method originating in \cite{Balasubramanian:2013lsa} 
and called ``hole-ography".
The hole-ography is a study about how a bulk curve and its circumference are holographically identified
by boundary entanglement entropy \cite{Balasubramanian:2013lsa}, 
and also provides a holographic definition of points and distances \cite{Czech:2014ppa},
by taking a shrinking limit of bulk curves.
One of the recent progress in this direction is about kinematic space and seen in \cite{Czech:2015qta,
Czech:2015kbp,Czech:2019hdd}.
Although reading metrics from the holographic distances is difficult, 
and hole-ography alone cannot describe temporal components of metrics,
the holographic distances would be helpful to determine conformal factors.

In order to examine if the guess works, we restrict ourselves to locally AdS$_3$ static spacetimes,
where the hole-ography has been well-studied.
Since our concern is the unknown local conformal factor, each spacetime point requires only one non-null vector 
of which norm is holographically given.
Here we expect the hole-ography to provide such a non-null vector.
Though this naively appears possible always, the strategy is not so trivial,
since we do not know the correct one-to-one map between the set of holographic points 
by the light-cone cuts method and that by the hole-ography.
Finding such a map enables us to compute complete bulk metrics.

There is a situation where the map is easily found, the situation that entanglement wedge and causal wedge coincide.
Under the assumption of the two wedge coinciding,
we see that the second issue about finding cuts can also be carried out easily;
light-cone cuts are computed directly from entanglement entropy, through
the holographic definition of points in the hole-ography.
We will see the above strategy work well under the assumption,
and typical geometries, pure AdS$_3$, AdS$_3$ soliton, and BTZ black hole are successfully reconstructed.
It is noteworthy that BTZ black hole can be reconstructed, because the bulk-point singularity fails to find 
light-cone cuts in this geometry, as mentioned in \cite{Engelhardt:2016crc}.
 
Although our reconstruction method relies on the assumption, we expect the present study to lead to 
the discovery of more generic solutions for the two issues about the light-cone cuts method.
In addition, we reveal that there is a relation between correlator divergence 
and entanglement entropy, for light-cone cuts are identified by entanglement entropy in our method.
Also, in light of our study, it turns out to be effective to combine the two methods in reconstructing bulk.

This paper is organized as follows.
In section \ref{sec: reviews}, we briefly review the light-cone cuts method and the hole-ography.
In section \ref{sec: reconstruction}, we see how to obtain light-cone cuts from entanglement entropy
under the above assumption, and provide the reconstruction method with conformal factors determined
by the hole-ography.
Our method is applied to the three examples in section \ref{sec: examples}.
Section \ref{sec: discussion} is devoted for the summary, and discussions of possible extensions.
In appendix \ref{app: by geodesics}, we present another approach to determine conformal factors
by using the geodesic equation, instead of using the holographic distances.

\section{Review of the two methods}\label{sec: reviews}
We briefly review the light-cone cuts method \cite{Engelhardt:2016wgb} and 
the hole-ography about points and distances \cite{Czech:2014ppa} in this section.

\begin{figure}[t]
	\centering
	\begin{minipage}{0.45\columnwidth}
		\centering
		(a)~~~~~\\
		\includegraphics[height = 5cm]{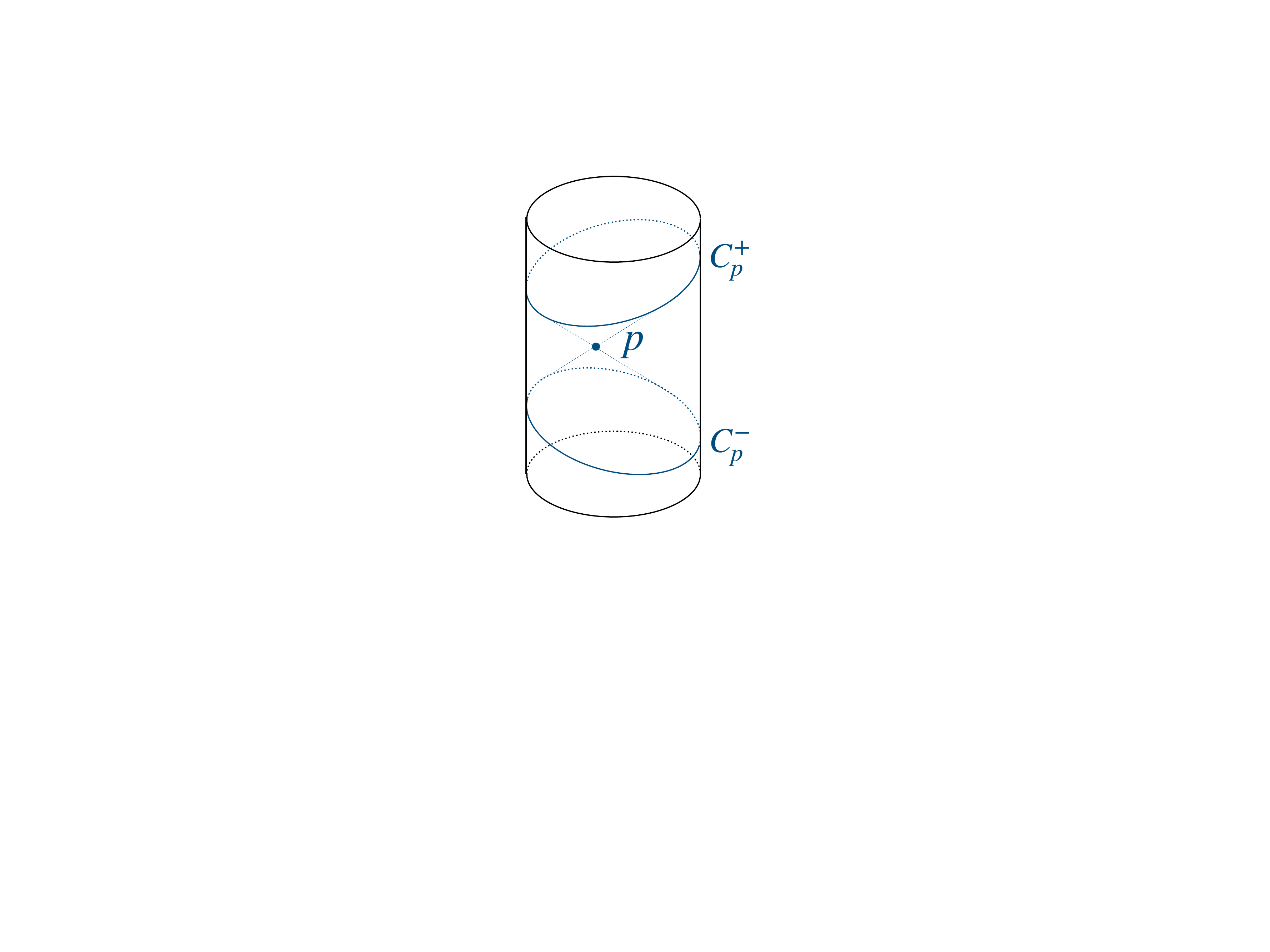}
	\end{minipage}
	\begin{minipage}{0.45\columnwidth}
		\centering
		(b)\\
		\includegraphics[height = 5cm]{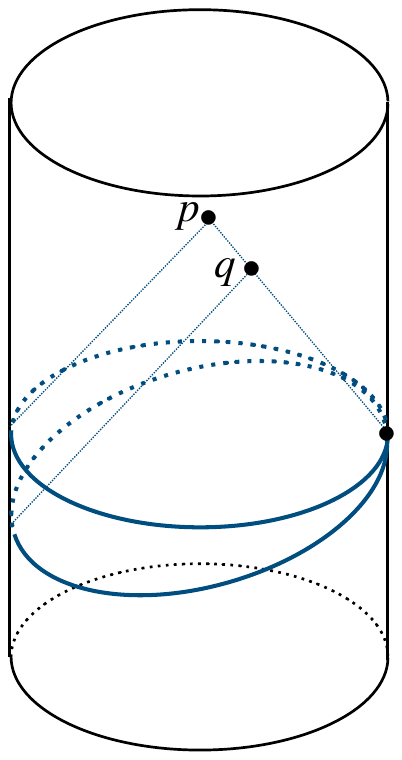}
\end{minipage}
\caption{(a) Future and past light-cone cuts
(b) Null-separated points and corresponding cuts.}
\label{fig: cuts}
\end{figure}

\subsection{Light-cone cuts}\label{subsec: cuts}
Let $M$ be the asymptotically AdS spacetime,\footnote{
In the original paper \cite{Engelhardt:2016wgb},
 more technical assumptions are imposed on $M$ to prove several facts introduced below.}
and $J^-(p)$ be the causal past of $p\in M$.
The past light-cone cut (or in short, past cut) of $p\in M$, $C^-_p$, is defined as 
\begin{align}
	C^-_p := \partial J^-(p)\cap \partial M,
	\label{eq: def of cuts}
\end{align}
where $\partial$ denotes boundaries.
The future light-cone cut $C^+_p$ is similarly defined, and those two cuts of $p$ are depicted in Fig.\ref{fig: cuts}(a).

The following properties of past cuts are derived from the causal structure of the bulk and boundary.
(Though written for past cuts, the following statements are valid for future cuts with
$+$ and $-$ subscripts exchanged.)
\begin{itemize}
	\item Each past cut is a complete spacelike hypersurface on $\partial M$.
	\item If $p\in I^+(\partial M)$, there exists a unique past cut $C^-_p$, where $I^+$ denotes timelike future.
	\item The intersection $C_p^-\cap C_q^-$ has non-empty open set, if and only if $p = q$.
\end{itemize}
These properties ensure a one-to-one correspondence between the set of past cuts and $I^+(\partial M)$.
Furthermore, the third property says that cuts can be distinguishable only with partial shapes of cuts.
The important property which is crucial for determining metrics is the following one (Fig.\ref{fig: cuts}(b)).
\begin{itemize}
	\item If $C^-_p$ and $C^-_q$ intersect at precisely one point, where both cuts are smooth, then
		$p$ and $q$ are null-separated.
\end{itemize}

The above statements are causal information cuts can describe,
which can be shown only by investigating causal structure --- no holographic interpretation has been used yet.
The set of cuts is of course different depending on the bulk geometry, and the information of the bulk geometry is
expected to be encoded in some boundary observable.
If cuts are identified from boundary theory, it seems helpful to use cuts in reconstructing bulk geometries.

It is indicated in \cite{Engelhardt:2016wgb} that correlation functions have data of light-cone cuts.
In general, a correlator diverges when all points are null-separated from its interaction point.
It is shown in \cite{Maldacena:2015iua} that there is a class of correlator divergences caused by
such interaction points being in the bulk.
Though the analysis there is restricted to the dual theories of low dimensional pure AdS,
it is discussed in \cite{Engelhardt:2016wgb} that there must exist such divergences also
in other asymptotically AdS spacetimes.
Using the bulk-point singularity is appealing, for it implies that cuts can be obtained from boundary theory in principle.
However we have to find correlators having five or more points, and hence is difficult.
Thus, we do not use it in the reconstruction method introduced later, hence we skip reviewing this part.

Next, we see how to use cuts to describe bulk geometries.
The set of cuts completely determines the bulk causal structure,
in other words, the metric up to an conformal rescaling, which is called conformal metric.
Let us suppose that cuts are obtained from the boundary theory.\footnote{
Since an open subset of a cut is enough for identifying the cut,
we do not always have to collect the whole shape of each cut.
}
Because there exists in principle a one-to-one correspondence between the set of past (future) cuts and 
$\partial I^+(\partial M)$ ($\partial I^-(\partial M)$),
cuts must be distinguished by an index space $\Lambda = \{\lambda\}$ having the same dimension as the bulk, $d$.
Our holographic interpretation is that the index space is the set of bulk points,\footnote{
What bulk region the index space describes depends on how cuts are obtained.
In our method introduced later, this matter is governed by the hole-ography.
} and $\lambda$
is a bulk point described in some coordinate.

Let us write each cuts obtained from the boundary theory as $C^\pm_\lambda$.\footnote{
In this paper, $p,q,\cdots$ are used to denote bulk points in bulk analysis, while
$\lambda$ is used to denote bulk points holographically defined from the boundary theory. 
}
Recalling the fourth property of cuts above, if $C^-_\lambda$ and $C^-_{\lambda+\delta\lambda}$
are tangent at one point, $\delta\lambda$ must be a null vector at $\lambda\in \Lambda$.
If we take $d(d+1)/2$ of such vectors at $\lambda$ and write them as $n_i(\lambda)$,
the metric at $\lambda$ is determined by solving simultaneous equation
\begin{align}
	g_{\mu\nu}(\lambda) n_i^\mu(\lambda)n_i^\nu(\lambda) = 0\qquad (i = 1,\cdots, d(d+1)/2).
\end{align}
However, the conformal factor cannot be obtained, since this equation is homogeneous,
in short, the conformal metric is reconstructed.

\subsection{Hole-ography}\label{subsec: hole-ography}
The hole-ography is originally a method to measure circumferences of bulk curves \cite{Balasubramanian:2013lsa},
and later used to give holographic definition of bulk points and distances on each time slice \cite{Czech:2014ppa}.
Here we review \cite{Czech:2014ppa}.

\begin{figure}[t]
	\centering
	\begin{minipage}{0.45\columnwidth}
	\centering
		(a)\\
		\includegraphics[height = 4cm]{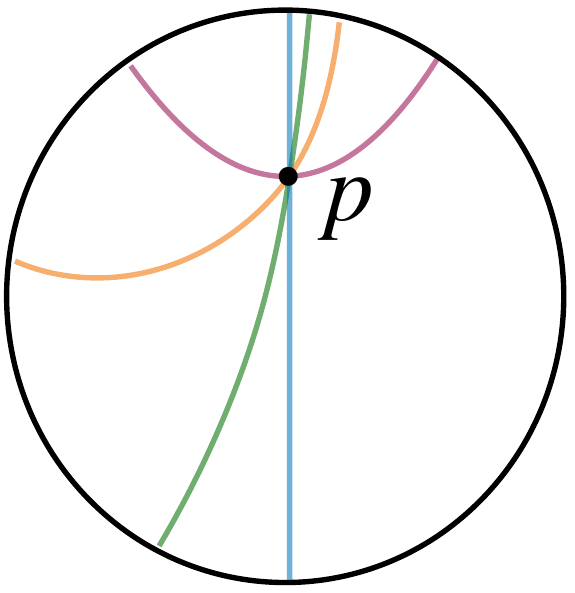}
	\end{minipage}
	\begin{minipage}{0.45\columnwidth}
		\centering
		(b)~~~~~~~~\\
		\includegraphics[height = 4cm]{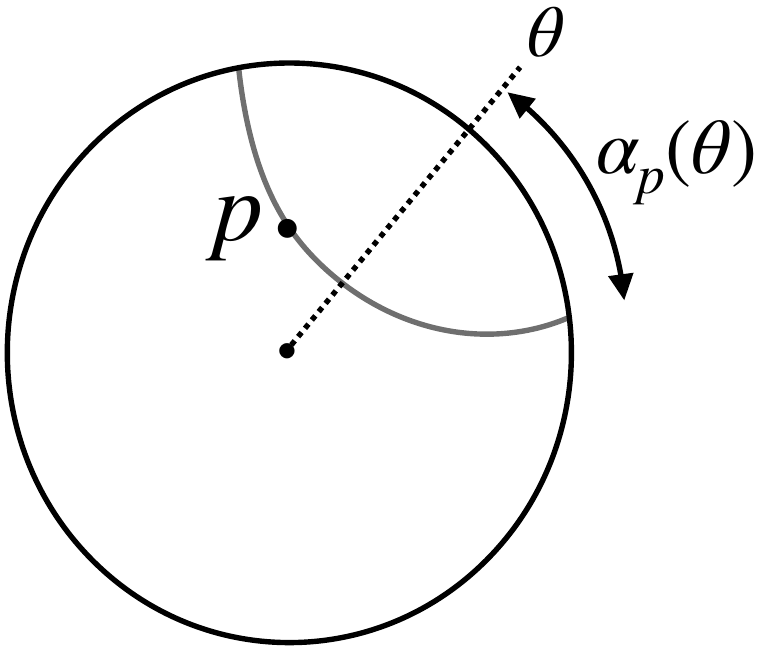}
	\end{minipage}
	\caption{(a) A family of geodesics (b) The boundary interval in $\theta$-direction}
	\label{fig: boundary interval}
\end{figure}

Let us consider an asymptotically AdS$_3$ static spacetime and suppose that the boundary metric is 
$\d s^2 = -\d t^2 + L^2\d \theta^2$.
For a point $p$ on a time slice $\Sigma$, there exists a family of inextensible spatial geodesics on $\Sigma$
which pass through $p$ (Fig.\ref{fig: boundary interval}(a)).
The two ends of each geodesic are on the boundary, and the angular interval between the two ends is
expressed as $I_p(\theta) = [\theta - \alpha_p(\theta),\theta + \alpha_p(\theta)]$, where
$\theta$ is the center of the interval (Fig.\ref{fig: boundary interval}(b)).\footnote{
In general, $\alpha_p$ can be multivalued, which is solved to be single-valued by taking the covering of $\theta$.}
We call $\alpha_p$ \textit{point function} in this paper.\footnote{
The name of $\alpha_p$ differs in each literature.\label{foot: terminology}
}
There must exist a one-to-one correspondence between the set of point functions and set of points on $\Sigma$.
In addition, each point function has the information about spatial geodesics, and hence about 
geodesic distances.
Thus, if point functions are derived from the boundary theory, it can give a holographic
definition of points and distances on $\Sigma$.

Let $S(\alpha)$ be the entanglement entropy of boundary interval with length $2\alpha$.
Here we have assumed the translational symmetry about $\theta$ on the boundary, so $S(\alpha)$
depends only on the length of the interval.
It is conjectured in \cite{Czech:2014ppa} that $\alpha$ is a point function if it extremizes
\begin{align}
	I[\alpha] = \int\d\theta\sqrt{-S''(\alpha(\theta))(1-\alpha'(\theta)^2)},
	\label{eq: action}
\end{align}
of which Euler-Lagrange equation is
\begin{align}
	(1-\alpha'(\theta)^2)S'''(\alpha(\theta)) + 2\alpha''(\theta)S''(\alpha(\theta)) = 0.
	\label{eq: EL equation}
\end{align}
Here, $S'(\alpha(\theta))$ is the value obtained by substituting $\alpha(\theta)$ into derivative function $S'(\alpha)$.
Since \eqref{eq: EL equation} is a second order differential equation, $\alpha$ acquires two integral constants.
Combining them with the time of $\Sigma$, we regard the list of the three constants as a bulk point outside the
horizon.

This conjecture is derived by analyzing the geometry of pure AdS$_3$,
and confirmed for conical defect AdS$_3$ and BTZ black hole.
However, it is pointed out in \cite{Burda:2018rpb} that \eqref{eq: EL equation} fails for BTZ bubble spacetime,
where $S(\alpha)$ is computed from Ryu-Takayanagi formula, assuming the bulk geometry at first.
Though we do not know if BTZ bubble has a dual field theory where Ryu-Takayanagi formula holds,
\eqref{eq: action} might be forced to be modified, or at least, should be modified if one wants to use it in more general cases.

Let us adopt their conjecture here and consider how distances on $\Sigma$ can be measured from the boundary.
We use $\alpha_\lambda$ to express point functions, where $\lambda^0$ is the time of $\Sigma$,
and $\lambda^1$ and $\lambda^2$ are integral constants of \eqref{eq: EL equation}; 
$\lambda = (\lambda^0,\lambda^1,\lambda^2)$ is regarded as a bulk point.
Again in \cite{Czech:2014ppa}, they showed the following statement for pure AdS$_3$, conical defect AdS$_3$,
and BTZ black hole (the outside of the horizon).
\begin{itemize}
	\item Let $\alpha_{\lambda_i}\,(i=1,2)$ be different point functions, and suppose that $\alpha_i$ is minimized
		at $\theta = \phi_i$.
		If there exists unique $\theta_\mathrm{int}$ between $\phi_1$ and $\phi_2$ such that 
		$\alpha_{\lambda_1}(\theta_\mathrm{int}) = \alpha_{\lambda_2}(\theta_\mathrm{int})$, 
		the geodesic distance between $\lambda_1$ and $\lambda_2$ is given by
		\begin{align}
			d(\lambda_1,\lambda_2) = 2G\l|
				\int_{\phi_1}^{\theta_\mathrm{int}}\d\theta\, S'(\alpha_{\lambda_1}(\theta))
				+
				\int_{\theta_\mathrm{int}}^{\phi_2}\d\theta\,S'(\alpha_{\lambda_2}(\theta))
			\r|.
			\label{eq: distance}
		\end{align}
\end{itemize}
Following the analysis of the above three examples in \cite{Czech:2014ppa},
we can accept that \eqref{eq: distance} would work for bulk geometries invariant under $\theta$-translation,
which from the boundary perspective corresponds to the field theory being in a state invariant under the same translation.
Now we are focusing on the case that entanglement entropy depends only on the interval length, 
it is reasonable to consider \eqref{eq: distance} to be valid.
Thus, we will adopt \eqref{eq: distance} in the reconstruction method introduced later.
(There would be many alternative tools within the hole-ography to accomplish the reconstruction,
one of which is shown in appendix \ref{app: by geodesics}.)


\section{Reconstruction of bulk geometries}\label{sec: reconstruction}
In the previous section, we have seen that light-cone cuts cannot determine conformal factors,
and the hole-ography cannot determine temporal components.
Generic answers to overcome these disadvantages are still under investigation.
In order to make progress in those problems, we solve these problems for theories such that 
causal and entanglement wedge coincide.
We first see how light-cone cuts are related to the hole-ography under the coincidence.
Secondly in subsection \ref{subsec: conformal factor}, we give a formula of infinitesimal distances 
by entanglement entropy, by which conformal factors are identified.
Finally in subsection \ref{subsec: general procedure}, our reconstruction procedure is proposed.

\subsection{Connection between light-cone cuts and hole-ography}
\begin{figure}[t]
	\centering
	\begin{minipage}{0.45\columnwidth}
		(a)\\
		\centering
		\includegraphics[height = 5cm]{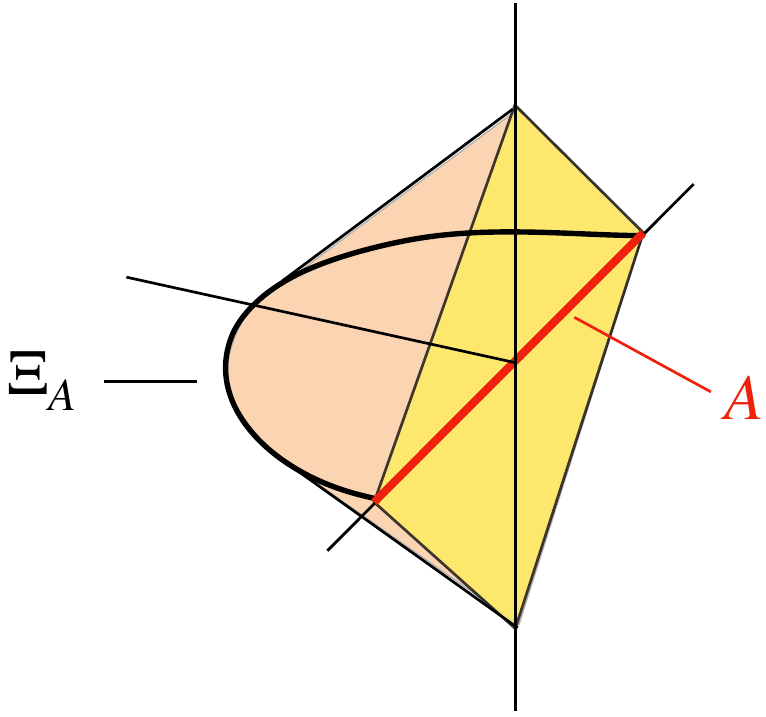}
	\end{minipage}
	\begin{minipage}{0.45\columnwidth}
		(b)\\
		\centering
		\includegraphics[height = 5cm]{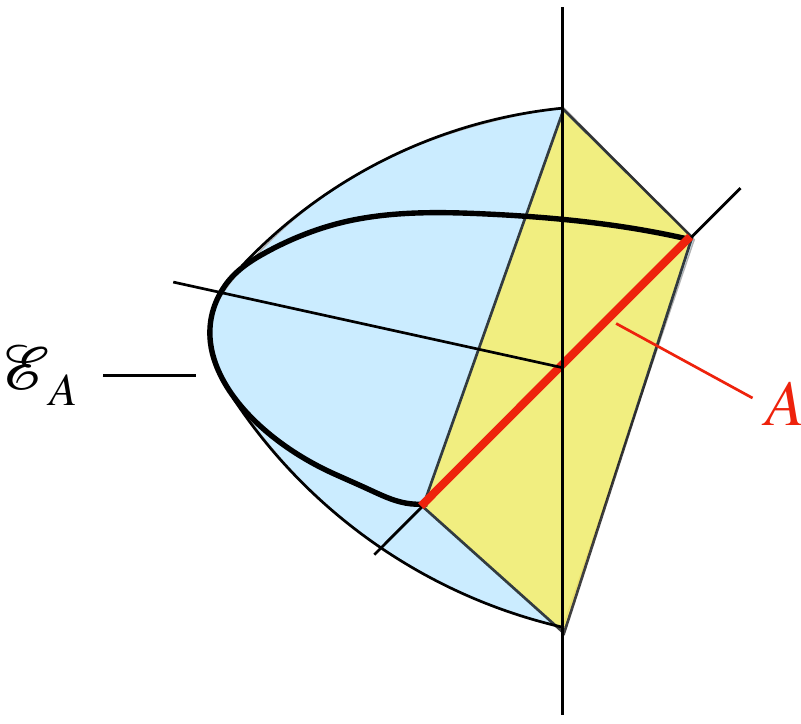}
	\end{minipage}	
	\caption{(a) Causal wedge (b)Entanglement wedge}
	\label{fig: wedges}
\end{figure}

In this subsection, we see that light-cone cuts can be identified as point functions,\footnote{
The explanation about this terminology is in footnote \ref{foot: terminology}.}
when causal and entanglement wedge coincide.

Let us first confirm the definition and properties of the two wedges.
We consider the case that the bulk is static and its dimension is three, 
where the reconstruction method introduced later is valid.
Let $A$ be a boundary interval on time slice $\Sigma$, and $\Diamond_A$ be the boundary domain of dependence of $A$.
The \textit{causal wedge} of $A$ is defined as the bulk region 
$J^-(\Diamond_A )\cap J^+(\Diamond_A)$ (Fig.\ref{fig: wedges}(a)).
We write the one-dimensional bifurcation surface as $\Xi_A$.
The minimal surface\footnote{
For example in a geometry with a black hole, we focus on the case where $A$ is not so large, 
hence the minimal surface becomes a geodesic.
This causes no problem to our reconstruction method (see the last paragraph of subsection \ref{subsec: soliton}).
}
of $A$, $\mathcal E_A$, lies on $\Sigma$, and let $\mathcal R_A$ be the surface
on $\Sigma$ with $\partial \mathcal R_A = A\cup \mathcal E_A$.
The \textit{entanglement wedge} of $A$ is the bulk domain of dependence of $\mathcal R_A$ (Fig.\ref{fig: wedges}(b)).

\begin{figure}[t]
	\centering
	\includegraphics[width = 6cm]{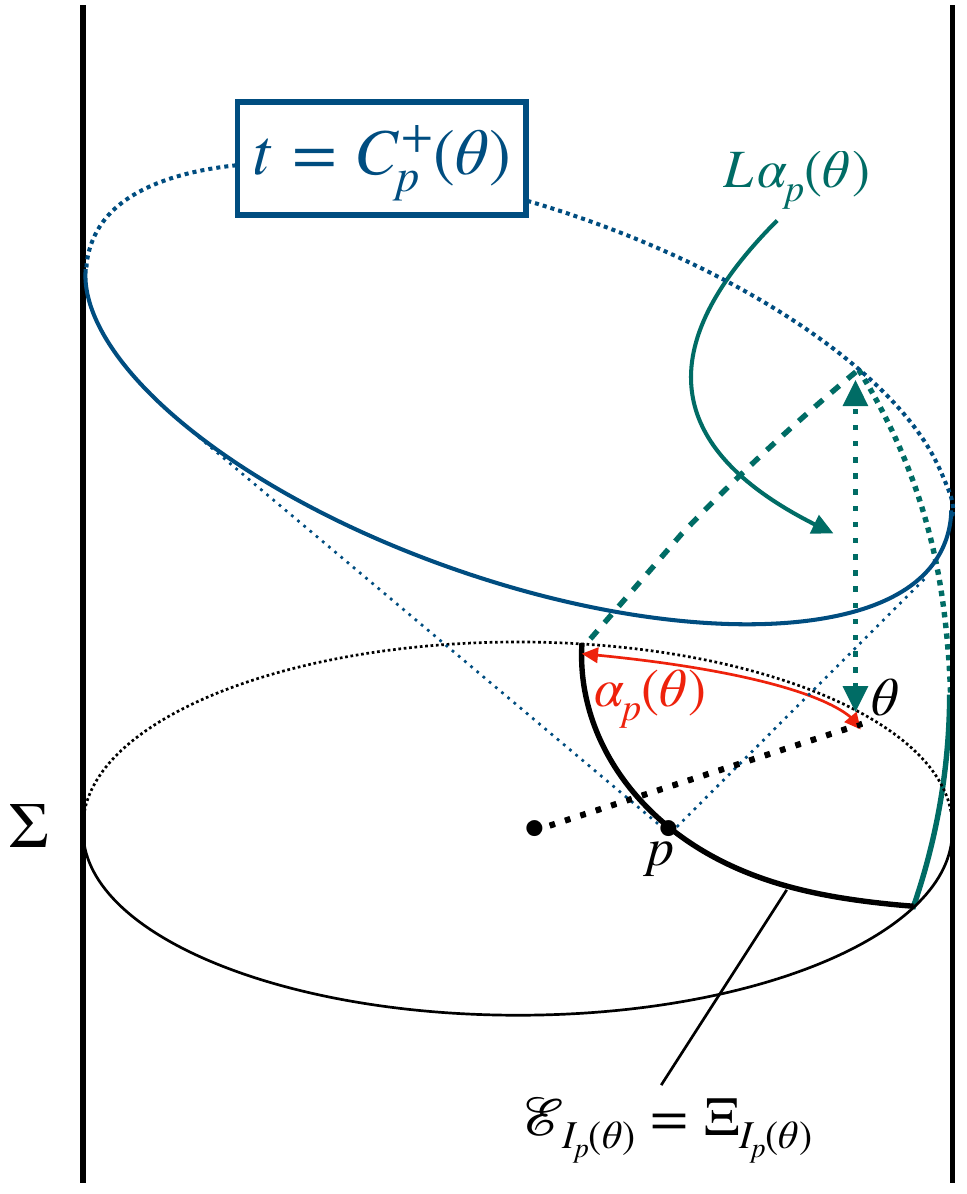}
	\caption{The future cut and point function of bulk point $p$}
	\label{fig: cut and point function}
\end{figure}

As having mentioned in section \ref{sec: introduction}, we consider condition $\Xi_A = \mathcal E_A$\footnote{
This assumption has been partially characterized in boundary language in \cite{Cardy:2016fqc}.
}
to find a simple relation between cuts and point functions.
To see the relation, we take a point function $\alpha_p$ and a related boundary interval 
$I_p(\theta) = [\theta-\alpha_p(\theta), \theta + \alpha_p(\theta)]$.
If $\Xi_{I_p(\theta)} = \mathcal E_{I_p(\theta)}$, both of the top and bottom vertex of $\Diamond_{I_p(\theta)}$ are
null-related to $p$, since $p$ is by definition on $\mathcal E_{I_p(\theta)}$, i.e., on $\Xi_{I_p(\theta)}$.
Thus, the top vertex is on $C_p^+$ and the bottom one is on $C^-_p$.
If we move $\theta$, the orbit of the top vertex is the future cut of $p$, and that of the bottom one is the past cut
(see Fig.\ref{fig: cut and point function}).
Therefore, the future and past cut of $p$, $C^\pm_p$, are written as
\begin{align}
	C^\pm_p(\theta) = t_0 \pm L\alpha_p(\theta),
	\label{eq: cut and point function}
\end{align}
with $t_0$ being the time of $\Sigma$.	
Conversely, if \eqref{eq: cut and point function} holds for any $p$, the two wedges coincide for any interval.

Equation \eqref{eq: cut and point function} implies that once we get point functions from boundary entanglement entropy
through \eqref{eq: EL equation}, they directly become cuts.
Thus, the conformal metric and spatial distances on each time slice are both encoded in entanglement entropy.
One would also think of the idea that we start with finding cuts by the bulk-point singularity, and then obtain
point functions through \eqref{eq: cut and point function}.
However, we are to use entanglement entropy to determine conformal factors in our method.

\subsection{Conformal factors by hole-ography}\label{subsec: conformal factor}
Let us suppose that the light-cone cuts method has already given the conformal metric of the bulk.
Now that our interest is how to determine the conformal factor, 
there is no need to examine spatial distances thoroughly.
In this subsection, we give a generic way to find and calculate a convenient distance at each point, by using \eqref{eq: distance}.
A way to determine conformal factors without \eqref{eq: distance} is shown 
in appendix \ref{app: by geodesics}.

Since we are supposing that we have followed the light-cone cuts method, 
the conformal metric is given for $\lambda$ labeling cuts as $C_\lambda^\pm$.
Assuming that entanglement and causal wedge coincide, we have \eqref{eq: cut and point function}, and hence
point functions are also labeled by $\lambda$.
It means that certain spatial distances are given in terms of $\lambda$ through \eqref{eq: distance}.
Therefore, if we find $\delta \lambda$ at $\lambda$ such that \eqref{eq: distance} can be applied,
condition $d^2(\lambda,\lambda+\delta\lambda) = g_{\mu\nu}(\lambda)\delta\lambda^\mu\delta\lambda^\nu$
determines the conformal factor at $\lambda$, where $g_{\mu\nu}$ is the reconstructed metric with its conformal factor
still unknown.

\begin{figure}[t]
	\centering
	\includegraphics[width = 9cm]{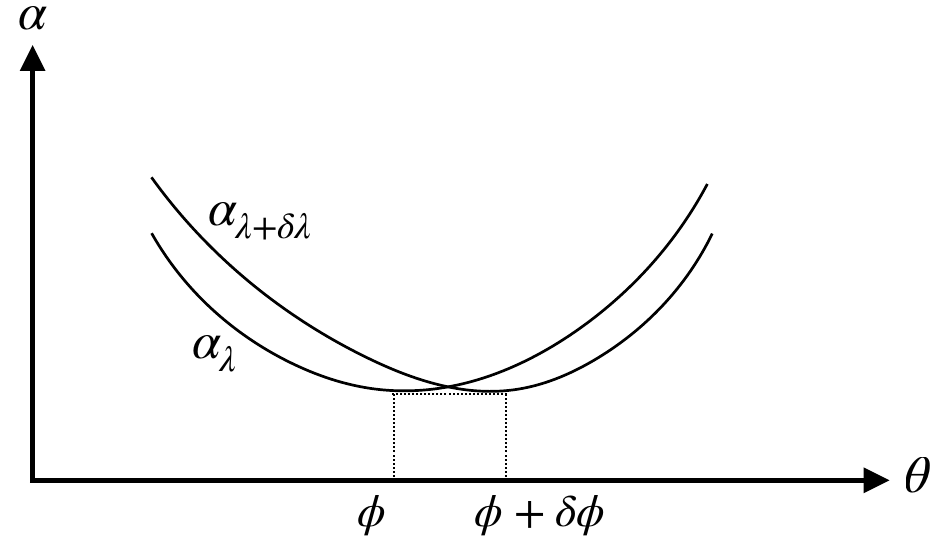}
	\caption{The intersection between two infinitesimally different point functions.}
	\label{fig: intersection}
\end{figure}

There is a way to find such $\delta\lambda~(\delta\lambda^0 = 0)$ for all $\lambda$.
Let $\phi$ be an angle minimizing $\alpha_\lambda$, i.e., $\alpha_\lambda'(\phi) = 0$.\footnote{
There must exist such $\phi$ because $\alpha_\lambda$ is interpreted as boundary intervals subtended by
geodesics passing through $\lambda$.
}
If $\alpha_{\lambda}''(\phi)\neq 0$ and there exists infinitesimal $\delta\phi$ such that
\begin{align}
	\alpha_{\lambda + \delta\lambda}'(\phi+\delta\phi) = 0\qquad
	\mathrm{and}
	\qquad
	\alpha_{\lambda+\delta\lambda}(\phi+\delta\phi) = \alpha_\lambda(\phi),
	\label{eq: intersection}
\end{align}
then $\alpha_{\lambda}$ and $\alpha_{\lambda+\delta\lambda}$ intersect once between
$\phi$ and $\phi+\delta\phi$ (see Fig.\ref{fig: intersection}).
Expanding \eqref{eq: intersection} to the first order of $\delta$, we obtain
\begin{align}
	\delta\lambda^i\frac{\partial\alpha_\lambda}{\partial\lambda^i}(\phi) = 0,\qquad
	\delta\phi = - \frac{\delta\lambda^i}{\alpha_\lambda''(\phi)}\frac{\partial\alpha_\lambda'}{\partial\lambda^i}(\phi),
	\label{eq: delta determined}
\end{align}
with $i$ contracted to run over $\{1,2\}$.
We choose $\delta\lambda$ from the former equation and $\delta\phi$ from the latter.
Finally by using \eqref{eq: distance}, the distance between $\lambda$ and $\lambda + \delta\lambda$ is calculated
to $O(\delta\lambda^1)$ as
\begin{align}
	d(\lambda,\lambda+\delta\lambda) = 2G\Bigl|\delta\phi\,S'(\alpha_\lambda(\phi))\Bigr|
\end{align}
with $\delta\phi$ given in \eqref{eq: delta determined}.
All we have to do is to compare this with $g_{\mu\nu}\delta\lambda^\mu\delta\lambda^\nu$ given by the conformal metric.

There could exist other ways to calculate an infinitesimal distance at $\lambda$.
Though we can expect that the way introduced here is valid in most cases where the hole-ography works,
we have to find another way if $\alpha''_\lambda(\phi) = 0$.

\subsection{Reconstruction procedure}\label{subsec: general procedure}
Now we are ready to propose our reconstruction procedure.
We assume that the bulk is static, locally AdS$_3$ in the global patch, and invariant under the rotation ($\theta$-translation).
In terms of the boundary language, the coordinate is $(t,\theta)$ with $\d s^2 = -\d t^2 + L^2\d\theta^2$, and
the field theory is in a $t$- and $\theta$-independent state.

Our reconstruction procedure is as follows.
All statements are of course written in terms of the boundary language.
\begin{enumerate}
\renewcommand{\labelenumi}{(\arabic{enumi})}
	\item \label{axiom: point function}
		Let us take a time slice $t = \lambda^0$ on the boundary.
		We call $\alpha$ \textit{point function} on $t = \lambda^0$, if a function extremizes \eqref{eq: action},
		\begin{align}
			I[\alpha] = \int\d\theta\sqrt{-S''(\alpha(\theta))(1-\alpha'(\theta)^2)}.
		\end{align}
		The Euler-Lagrange equation is given as \eqref{eq: EL equation},
		\begin{align}
			(1-\alpha'(\theta)^2)S'''(\alpha(\theta)) + 2\alpha''(\theta)S''(\alpha(\theta)) = 0.
			\label{eq: re EL equation}
		\end{align}
	\item \label{axiom: points}
		The Euler-Lagrange equation being solved, $\alpha$ acquires two integral constants $\lambda^1$ and $\lambda^2$,
		and is labeled as $\alpha_\lambda$ with $\lambda = (\lambda^0,\lambda^1,\lambda^2)$.
		We consider the set of $\lambda$, $\Lambda = \{\lambda\}$, to be the set of bulk points outside the horizon, 
		and the difference in how to select integral constants corresponds to the coordinate choice of the bulk.
	\item \label{axiom: cuts}
		The set of past cuts\footnote{
		Future cuts are also used instead of past cuts.
		} is identified as
		\begin{align}
			\{C^-_\lambda = \lambda^0 - L\alpha_\lambda\,|\,\lambda \in \Lambda \}.
		\end{align}
	\item \label{axiom: conformal metric}
		If $C_{\lambda}^-$ and $C_{\lambda + \delta\lambda}^-$ are tangent at one point, $\delta\lambda$
		is an infinitesimal null vector at $\lambda\in \Lambda$.
		We find six such $\delta\lambda$'s, which we write as $n_i(\lambda)\,(i=1,\cdots,6)$, 
		and solve the following simultaneous equation for $g_{\mu\nu}(\lambda)$:
		\begin{align}
			0 = g_{\mu\nu}(\lambda) n_i^\mu(\lambda) n_i^\nu (\lambda) \qquad(i=1,\cdots,6).
		\end{align}
		Then we obtain the conformal metric.
		The conformal factor cannot be determined, since this equation is homogeneous.
	\item  \label{axiom: distances}
		Let $\phi$ be the angle minimizing $\alpha_\lambda$, and $\phi + \delta\phi$ be the one minimizing 
		$\alpha_{\lambda+\delta\lambda}~(\delta\lambda^0 = 0)$.
		If there exists unique $\theta_\mathrm{int}$ between $\phi$ and $\phi + \delta\phi$ such that 
		$\alpha_{\lambda}(\theta_\mathrm{int}) = \alpha_{\lambda + \delta\lambda}(\theta_\mathrm{int})$, 
		the distance between $\lambda$ and $\lambda + \delta\lambda$ is given by
		\begin{align}
			d(\lambda,\lambda + \delta\lambda) = 2G\l|
				\int_{\phi}^{\theta_\mathrm{int}}\d\theta\, S'(\alpha_{\lambda}(\theta))
				+
				\int_{\theta_\mathrm{int}}^{\phi + \delta\phi}\d\theta\,S'(\alpha_{\lambda + \delta\lambda}(\theta))
			\r|.
			\label{eq: re distance}
		\end{align}
		 This formula is reduced to
		\begin{align}
			d(\lambda,\lambda+\delta\lambda) = 2G\Bigl|\delta\phi\,S'(\alpha_\lambda(\phi))\Bigr|,	
		\end{align}
		if $\alpha_\lambda''(\phi)\neq 0 $ and $\delta\lambda$ and $\delta\phi$ are taken as
		\begin{align}
			\delta\lambda^i\frac{\partial\alpha_\lambda}{\partial\lambda^i}(\phi) = 0,\qquad
			\delta\phi = - \frac{\delta\lambda^i}{\alpha_\lambda''(\phi)}\frac{\partial\alpha_\lambda'}{\partial\lambda^i}(\phi).
			\label{eq: re intersection}
		\end{align}
	\item \label{axiom: conformal factor}
		Finally, the conformal factor is determined by 
		\begin{align}
			2G\Bigl|\delta\phi\,S'(\alpha_\lambda(\phi))\Bigr| = \sqrt{g_{\mu\nu}(\lambda)\delta\lambda^\mu\delta\lambda^\nu}\,.
			\label{eq: comparison}
		\end{align}
\end{enumerate}

We have to confirm if the two wedges actually coincide, purely within boundary theory at the process \eqref{axiom: cuts}.
This can be confirmed as follows:
we arbitrarily take two boundary points on $C^-_\lambda$ and one point on $C^+_\lambda$,
then confirm if there exists a point on $C^+_\lambda$ such that some boundary correlator
of the four point diverges.
If such a point exists, the divergence is the bulk-point singularity, and hence $C^\pm_\lambda$
is the true cuts.\footnote{
However, the confirmation method does not work well in black hole geometries \cite{Engelhardt:2016crc},
and as well in AdS soliton.
}

One might have the idea that there is no need to assume the coincidence of the two wedges; 
it is naively possible that we identify cuts from the bulk-point singularity to obtain the conformal metric, and
determine the conformal factor by following \eqref{axiom: distances} and \eqref{axiom: conformal factor}.
However in this case, cuts and point functions have different labels, in other words, they describe bulk points by
different coordinates.
Hence we cannot read the conformal factor from \eqref{eq: comparison}.
Unfortunately, for the time being, we do not know the coordinate transformation between them.

\section{Examples of explicit reconstruction}\label{sec: examples}
We reconstruct the three geometries, pure AdS$_3$, AdS$_3$ soliton, and BTZ black hole (outside the horizon),
by following the reconstruction procedure introduced in subsection \ref{subsec: general procedure}.

\subsection{Pure AdS$_3$}\label{subsec: AdS}
In vacuum CFT$_2$, the entanglement entropy of any interval with length $2\alpha$ is computed as
\begin{align}
	S(\alpha) = \frac{c}{3}\ln \l(\frac{2L}{\epsilon}\sin\alpha\r),
\end{align}
where $c$ is the central charge, $\epsilon$ is a UV regulator \cite{Calabrese:2004eu}.
The Euler-Lagrange equation \eqref{eq: re EL equation} is now reduced to
\begin{align}
	\alpha'(\theta)^2 - 1 + \alpha''(\theta)\tan\alpha(\theta) = 0,
\end{align}
which is solved to give
\begin{align}
	\alpha_{\lambda}(\theta) = \cos^{-1}[\lambda^1\cos(\theta - \lambda^2)],
	\label{eq: AdS pf}
\end{align}
by putting $\beta(\theta) = \cos\alpha(\theta)$.
Here $\lambda^1$ and $\lambda^2$ are integral constants.
Combining the time $\lambda^0$ and $(\lambda^1,\lambda^2)$, we regard $\lambda = (\lambda^0,\lambda^1,\lambda^2)$
as bulk points (procedure \eqref{axiom: points}).

Having obtained point functions, we immediately get past light-cone cuts from procedure \eqref{axiom: cuts}:
\begin{align}
	C^-_\lambda(\theta) = \lambda^0 - L\cos^{-1}[\lambda^1\cos(\theta - \lambda^2)].
\end{align}
According to procedure \eqref{axiom: conformal metric}, we have to find vector $n(\lambda,\theta)$\footnote{
According to procedure \eqref{axiom: conformal factor}, we need six vectors, so all we have to do is
to find $n(\lambda,\theta)$ satisfying \eqref{eq: AdS tangent} for six different $\theta$'s.
However, in the present case and the two remaining examples in this section, 
we get $n(\lambda,\theta)$ for all $\theta$.
} satisfying
\begin{align}
	C_\lambda(\theta) = C_{\lambda + \epsilon n},\qquad
	C'_\lambda(\theta) = C'_{\lambda + \epsilon n}(\theta),
	\label{eq: AdS tangent}
\end{align}
where $\epsilon$ is an infinitesimal constant.
Solving these equations to $O(\epsilon^1)$ gives
\begin{align}
	n(\lambda,\theta)  \propto &~ L\lambda^1\sqrt{1-(\lambda^1)^2\cos^2(\theta-\lambda^2)}
	\frac{\partial}{\partial\lambda^0}
	\nonumber\\
	&+ 	\lambda^1((\lambda^1)^2-1)\cos(\theta - \lambda^2)\frac{\partial}{\partial\lambda^1}
	-\sin(\theta - \lambda^2)\frac{\partial}{\partial\lambda^2}.
\end{align}
The conformal metric is computed through
\begin{align}
	\forall\theta,\quad g_{\mu\nu}(\lambda)n^\mu(\lambda,\theta)n^\nu(\lambda,\theta) = 0,
	\label{eq: AdS conformal equation}
\end{align}
and the result is
\begin{align}
		\d s^2 = e^{\omega(\lambda)}\l[
	-\{1-(\lambda^1)^2\}(\d \lambda^0)^2
	+ L^2(\d\lambda^1)^2 
	+ L^2(\lambda^1)^2\{1-(\lambda^1)^2\}(\d\lambda^2)^2
	\r],
	\label{eq: AdS conformal metric}
\end{align}
where $e^{\omega(\lambda)}$ is the undetermined conformal factor.

In order to determine $\omega$, let us calculate a spatial distance in a convenient direction, 
by following procedure \eqref{axiom: distances}.
The angle minimizing $\alpha_\lambda$ is obviously $\phi = \lambda^2$, and we see that 
$\delta\lambda$ and $\delta\phi$ related as $\delta\lambda^1 = 0$, $\delta\phi = \delta\lambda^2$
 satisfy \eqref{eq: re intersection}.
In this case, \eqref{eq: comparison} becomes
\begin{align}
	\frac{2Gc}{3}\frac{\lambda^1\delta\lambda^2}{\sqrt{1-(\lambda^1)^2}} 
	= L e^{\omega(\lambda)/2} \lambda^1\sqrt{1-(\lambda^1)^2}\,\delta\lambda^2,\quad
	\mathrm{i.e.}
	\quad
	e^{\omega(\lambda)} = \frac{1}{(1-(\lambda^1)^2)^2}.
\end{align}
Here we have used the well-known relation $c = 3L/2G$.
Thus, we conclude
\begin{align}
		\d s^2 = 
	-\frac{(\d \lambda^0)^2}{1-(\lambda^1)^2}
	+ \frac{(L\,\d\lambda^1)^2 }{\{1-(\lambda^1)^2\}^2}
	+ \frac{(L\,\lambda^1\d\lambda^2)^2}{1-(\lambda^1)^2},
\end{align}
which becomes the expression in the static coordinate
by coordinate transformation $\lambda^1\to \lambda^1/\sqrt{L^2+(\lambda^1)^2}$\,:
\begin{align}
		\d s^2 = -\l(1 + \frac{(\lambda^1)^2}{L^2}\r)(\d \lambda^0)^2 
	+ \frac{(\d\lambda^1)^2}{1 + (\lambda^1)^2/L^2} + (\lambda^1\d\lambda^2)^2.
	\label{eq: AdS final}
\end{align}
Now the pure AdS$_3$ geometry is reconstructed.

\subsection{AdS$_3$ soliton}\label{subsec: soliton}
The holographic entanglement entropy of AdS$_3$ geometry \eqref{eq: AdS soliton geometry}
 is given as
\begin{align}
	S(\alpha) = \frac{L}{2G}\ln\l[\frac{2r_0}{\epsilon}\sin\l(\frac{\alpha r_0}{L}\r)\r],
	\label{eq: soliton entanglement}
\end{align}
where $r_0$ is a constant length.
Calculations of holographic entanglement entropy are for example seen in \cite{Czech:2014ppa,Burda:2018rpb}.
Forgetting the bulk, we just assume that the entanglement entropy of the dual theory in a certain state is 
now computed as \eqref{eq: soliton entanglement}, and check if AdS$_3$ geometry is consistently reconstructed.

Euler-Lagrange equation \eqref{eq: re EL equation} is now 
\begin{align}
	\alpha'(\theta)^2-1 + \frac{L\alpha''(\theta)}{r_0}\tan\l(\frac{r_0\alpha(\theta)}{L}\r)=0,
\end{align}
and we solve this to obtain
\begin{align}
	\alpha_\lambda(\theta) = \frac{L}{r_0}\cos^{-1}\l[\lambda^1\cos\l(\frac{r_0}{L}(\theta-\lambda^2)\r)\r].
	\label{eq: soliton pf}
\end{align}
We regard $\{\lambda = (\lambda^0,\lambda^1,\lambda^2)\}$ as the set of bulk points with $\lambda^0$ denoting
the time coordinate.

Following procedure \eqref{axiom: cuts}, we get past cuts as
\begin{align}
	C^-_\lambda(\theta) = \lambda^0 - \frac{L^2}{r_0}\cos^{-1}\l[\lambda^1\cos\l(\frac{r_0}{L}(\theta-\lambda^2)\r)\r].
	\label{eq: soliton cut}
\end{align}
We find null vectors from \eqref{eq: AdS tangent} and the result is
\begin{align}
	n(\lambda,\theta)\propto &~
	L\lambda^1\sqrt{1-(\lambda^1)^2\cos^2(\theta-\lambda^2)}\frac{\partial}{\partial\lambda^0}\nonumber\\
	&+
	\frac{r_0}{L}\lambda^1((\lambda^1)^2-1)\cos\l(\frac{r_0}{L}(\theta-\lambda^2)\r)\frac{\partial}{\partial\lambda^1}
	-
	\sin\l(\frac{r_0}{L}(\theta-\lambda^2)\r)\frac{\partial}{\partial\lambda^2}.
\end{align}
Substituting this to \eqref{eq: AdS conformal equation}, we obtain the conformal metric,
\begin{align}
	\d s^2 = e^{\omega(\lambda)}\l[
	-(1-(\lambda^1)^2)(\d\lambda^0)^2
	+\frac{L^4}{r_0^2}(\d\lambda^1)^2
	+L^2(\lambda^1)^2(1-(\lambda^1)^2)(\d\lambda^2)^2
	\r].
	\label{eq: soliton conformal}
\end{align}

The conformal factor $\omega$ is determined from procedure \eqref{axiom: distances} and \eqref{axiom: conformal factor}.
The angle minimizing $\alpha_\lambda$ is $\phi = \lambda^2$, and \eqref{eq: re intersection} is satisfied
by $\delta\lambda^1 = 0$ and $\delta\phi = \delta\lambda^2$.
Thus in this case, \eqref{eq: comparison} becomes
\begin{align}
	\frac{r_0 \lambda^1}{\sqrt{1-(\lambda^1)^2}}\,\delta\lambda^2 
	= e^{\omega(\lambda)/2}L\lambda^1\sqrt{1-(\lambda^1)^2}\,\delta\lambda^2,
	\quad
	\mathrm{i.e.}
	\quad
	e^{\omega(\lambda)} = \frac{(r_0/L)^2}{(1-(\lambda^1)^2)^2}.
\end{align}
Therefore, we have the complete metric
\begin{align}
	\d s^2 = -\frac{(r_0/L)^2(\d\lambda^0)^2}{1-(\lambda^1)^2}
	+ \frac{(L\,\d\lambda^1)^2}{(1-(\lambda^1)^2)^2}
	+ \frac{(r_0\,\lambda^1\d\lambda^2)^2}{1-(\lambda^1)^2},
\end{align}
which is reduced to the following form by coordinate transformation 
$\lambda^1\to \sqrt{(\lambda^1)^2-r_0^2}/\lambda^1$\,:
\begin{align}
	\d s^2 = -\l(\frac{\lambda^1}{L}\r)^2(\d\lambda^0)^2 
	+ \frac{(L\,\d\lambda^1)^2}{(\lambda^1)^2-r_0^2}
	+((\lambda^1)^2-r_0^2)(\d\lambda^2)^2.
	\label{eq: AdS soliton geometry}
\end{align}
This is exactly the geometry of AdS$_3$ soliton.

Although we have reproduced the exact metric, there is actually facts we have ignored.
Eq.\eqref{eq: soliton entanglement} is rather the entwinement than the entanglement entropy, in short, it is the length of the geodesic rather than that of the minimal surface.
In fact, one can check, through geodesic calculation, that \eqref{eq: soliton pf} characterizes $\lambda$ as the intersection point of the geodesic family, $\alpha_\lambda$.
Bulk analysis also reveals that \eqref{eq: soliton cut} is a set of points connected with $\lambda$ by null geodesics.
Among these points however, there are points which are timelike separated from $\lambda$, and according to the definition of \eqref{eq: def of cuts}, such points must be excluded for the light-cone cuts method to work.
But, since those excluded points correspond to somewhat large $|\theta-\lambda^2|$, our reconstruction is valid because solving \eqref{eq: AdS conformal equation} around $\theta \sim \lambda^2$ is sufficient to give \eqref{eq: soliton conformal}.
Nevertheless, \eqref{eq: AdS conformal equation} is satisfied by \eqref{eq: soliton conformal} for all $\theta$, which implies the possibility to use the ``false cuts," the cuts including the excluded points and related to null geodesics, instead of the ``true cuts."
The potential connection between the false cuts and the entwinement is discussed in section \ref{sec: discussion}.

\subsection{BTZ black hole}\label{subsec: BTZ}
We again start with the well-known holographic entanglement entropy of BTZ black hole,
\begin{align}
	S(\alpha) = \frac{L}{2G}\ln\l[\frac{2r_0}{\epsilon}\sinh\l(\frac{\alpha r_0}{L} \r) \r],
\end{align}
where $r_0$ is a constant length.
For this entropy, Euler-Lagrange equation \eqref{eq: re EL equation} is given as
\begin{align}
	\alpha'(\theta)^2-1 + \frac{L\alpha''(\theta)}{r_0}\tanh\l(\frac{r_0\alpha(\theta)}{L}\r)=0.
\end{align}
The solution of this equation is expressed as\footnote{
The same discussion in the last paragraph of the previous subsection is also applied to this case.
}
\begin{align}
	\alpha_\lambda(\theta) = \frac{L}{r_0}\cosh^{-1}\l[
	\lambda^1\cosh\l(\frac{r_0}{L}(\theta-\lambda^2) \r)
	\r].
	\label{eq: BTZ pf}
\end{align}

We obtain past cuts,
\begin{align}
	C^-_\lambda(\theta) = \lambda^0
	-
	\frac{L^2}{r_0}\cosh^{-1}\l[
	\lambda^1\cosh\l(\frac{r_0}{L}(\theta-\lambda^2) \r)\r],
	\label{eq: BTZ cut}
\end{align}
and imposing \eqref{eq: AdS tangent} gives null vectors at $\lambda$:
\begin{align}
	n(\lambda,\theta)\propto &~
	L\lambda^1\sqrt{(\lambda^1)^2\cosh^2(\theta-\lambda^2)-1}\frac{\partial}{\partial\lambda^0}\nonumber\\
	&+
	\frac{r_0}{L}\lambda^1((\lambda^1)^2-1)\cosh\l(\frac{r_0}{L}(\theta-\lambda^2)\r)\frac{\partial}{\partial\lambda^1}
	-
	\sinh\l(\frac{r_0}{L}(\theta-\lambda^2)\r)\frac{\partial}{\partial\lambda^2}.
\end{align}
This being substituted to \eqref{eq: AdS conformal equation}, the conformal metric is computed as
\begin{align}
	\d s^2 = e^{\omega(\lambda)}\l[
	-((\lambda^1)^2-1)(\d\lambda^0)^2
	+\frac{L^4}{r_0^2}(\d\lambda^1)^2
	+L^2(\lambda^1)^2((\lambda^1)^2-1)(\d\lambda^2)^2
	\r].
\end{align}

Let us follow procedure \eqref{axiom: distances} and \eqref{axiom: conformal factor}.
We again find that $\phi = \lambda^2$ and that \eqref{eq: re intersection} is satisfied by $\delta\lambda^1 =0$
and $\delta\phi = \delta\lambda^2$, hence \eqref{eq: comparison} becomes
\begin{align}
	\frac{r_0\lambda^1}{\sqrt{(\lambda^1)^2-1}}\,\delta\lambda^2
	=
	e^{\omega(\lambda)/2}L\lambda^1\sqrt{(\lambda^1)^2-1}\,\delta\lambda^2
	\quad
	\mathrm{i.e.}
	\quad
	e^{\omega(\lambda)} = \frac{(r_0/L)^2}{((\lambda^1)^2-1)^2}.
\end{align}
Thus, the complete metric is 
\begin{align}
	\d s^2 = -\frac{(r_0/L)^2(\d\lambda^0)^2}{(\lambda^1)^2-1}
	+ \frac{(L\,\d\lambda^1)^2}{((\lambda^1)^2-1)^2}
	+ \frac{(r_0\,\lambda^1\d\lambda^2)^2}{(\lambda^1)^2-1},
\end{align}
which is reduced to the following form by transformation $\lambda^1\to \lambda^1/\sqrt{(\lambda^1)^2 - r_0^2}$\,:
\begin{align}
	\d s^2 = -\frac{(\lambda^1)^2-r_0^2}{L^2}(\d\lambda^0)^2
	+\frac{(L\,\d\lambda^1)^2}{(\lambda^1)^2-r_0^2}
	+(\lambda^1\d \lambda^2)^2.
\end{align}
Now the geometry of BTZ black hole (outside the horizon) is reconstructed.

\section{Summary and discussions}\label{sec: discussion}
We have presented an approach for reconstructing locally AdS$_3$ static spacetimes, combining
the two method, light-cone cuts method and hole-ography.
When the field theory is in a state independent of $\theta$, the entanglement entropy only depends on 
the length of the interval, under which we can expect that the holographic definition of points in the hole-ography works.
We have considered a certain case that entanglement and causal wedge coincide, where there is a direct one-to-one
relation between light-cone cuts and point functions.
Thus, we can obtain cuts from entanglement entropy, then compute conformal metrics, and finally read conformal factors
from the holographic distances provided by the hole-ography.

Let us discuss future directions of this work.
The most crucial problem to solve is that our method is valid only under the assumption of the two wedges coinciding.
We naively wonder if something similar to our method can be established only within the light-cone cuts method
or only within the hole-ography.
In the former case, we should note that the set of cuts does not have the information about entanglement wedges.
This is because, from the bulk viewpoint, cuts are identified only by the causal structure, and the causal structure
cannot provide non-null surfaces.
In the latter case on the other hand, computing entanglement entropy for various kinds of regions 
in the hole-ography seems too hard in general.

Thus, we should use the right observable in the right situation in the reconstruction procedure.
Those things being considered, finding out how to match the set of holographic points
by cuts and that by the hole-ography might be a good way to improve our method.
In light of our study, requiring the coincidence of the two wedges picks up a certain conformal factor.
If we locally rescale the bulk metric by a conformal factor, then point functions change
while light-cone cuts remain unchanged, where the relation $C_\lambda^\pm = t_0 \pm L\alpha_\lambda$
is modified in dependence of the conformal factor.
Hence the question is now to find the modified relation in the bulk and interpret it to the boundary language.
We expect that studying in this direction also leads us to discover other ways to find cuts.
It would also helpful to investigate the relation between entanglement wedge and causal wedge, 
and such studies already exist  (see \cite{Hubeny:2012wa,Wall:2012uf} for example).

As another question, how deep into the bulk can cuts and point functions describe?
The only investigation of reconstructing the metric inside a black hole is proposed 
in \cite{Hashimoto:2021umd} by using complexity, and here we would like to discuss the possibility of 
doing it in our framework.
Let us first consider the potential of point functions.
In pure AdS$_3$, spacelike geodesics become minimal surfaces, and the minimal surface of each interval is
equal to that of the complement region.
This property should be reflected to point functions, and in fact \eqref{eq: AdS pf} has a symmetry 
$\alpha_\lambda(\theta + \pi) = \pi - \alpha_\lambda(\theta)$.
On the other hand, point functions of BTZ black hole \eqref{eq: BTZ pf} do not have the symmetry
(those of AdS$_3$ soliton \eqref{eq: soliton pf} as well).
This fact is of course a reflection of the property that the existence of a hole\footnote{
We call region $\lambda^1<r_0$ hole, which is empty in AdS$_3$ soliton, and a black hole in BTZ.}
makes the difference between the minimal surface of an interval and that of its complement,
which means that the boundary field theory is in a mixed state.
Thus, point functions appear to be subject to those constraints on extremal surfaces which are examined in
\cite{Hubeny:2012ry,Engelhardt:2013tra,Engelhardt:2015dta}.

However, there is an interesting feature of point functions.
Calculating geodesics, we see that \eqref{eq: soliton pf}\footnote{
By the calculation of geodesics, we find that \eqref{eq: soliton pf} should take the branch of $\cos^{-1}$
such that $\alpha_\lambda$ smoothly increases as $|\theta-\lambda^2|$ increases.
\label{foot: branch}}
and \eqref{eq: BTZ pf} reflect the information of geodesics rather than minimal surfaces,\footnote{
This is shown in \cite{Czech:2014ppa} for BTZ case, and the same behavior is observed
for AdS$_3$ soliton.
}
This is because a minimal surface wraps the hole exactly once when the corresponding interval is large,
while point functions include those geodesics going around the hole many times 
(see figure 14 of \cite{Czech:2014ppa}), of which corresponding $S(\alpha)$ is called entwinement 
\cite{Balasubramanian:2014sra}.
In addition, according to the analysis of \cite{Czech:2014ppa}, a certain continuation of point functions to
complex numbers makes it possible to include geodesics penetrating the horizon of BTZ black hole.
Thus, if we find a modified formulation of point functions by properly extended $S(\alpha)$, 
point functions might describe the inside of a black hole.
It should be emphasized that what makes cuts method not able to reconstruct the black hole interior
is using the bulk-point singularity, and the interior can be reconstructed if we have the complete set of cuts.\footnote{
Note that \eqref{eq: re EL equation} does not reproduce cuts corresponding to points inside the horizon.
}
In short, cuts derived from the modified point functions, which might be the ``false cuts" discussed in subsection \ref{subsec: soliton}, could reconstruct the inside of a black hole.

\subsection*{Acknowledgement}
I am grateful to Koji Hashimoto, for encouraging me to write this paper and giving various fruitful
comments on the draft.

\appendix
\section{Another way to obtain conformal factors}\label{app: by geodesics}
In this appendix, we propose another way to determine conformal factors, which also gives
the proper conformal factors for the examples dealt with in section \ref{sec: examples}.
Instead of using  \eqref{eq: re distance}, we focus on the fact that point functions have the information of 
spacelike geodesics, then the geodesic equation can be used to obtain conformal factors
up to rigid factors.
The remaining rigid factors are determined by Ryu-Takayanagi formula.
Although the proposal here does not require holographic distances \eqref{eq: re distance}, 
calculations become a little bit troublesome, and furthermore, 
this method is unlikely to include quantum effects unfortunately, for we use the geodesic equation.

From the bulk viewpoint, interval $I_p(\theta) = [\theta - \alpha_p(\theta),\theta+\alpha_p(\theta)]$ is subtended by
the geodesic which passes through $p$ and is centered at $\theta$.
If $\alpha_p(\theta) = \alpha_q(\theta)$ for a fixed $\theta$ with $p$ and $q$ on the same time slice, 
the above geodesic also passes through $q$.
This motivates us to propose the following holographic principle.
\begin{itemize}
	\item The set of holographic points 
		\begin{align}
			\{\lambda\,|\,\alpha_\lambda(\theta) = \mathrm{const}\,(\theta\mbox{: fixed}),~\lambda^0 = \mathrm{const}\}
		\end{align}
		forms a geodesic.
\end{itemize}
Thus, instead of procedure \eqref{axiom: distances} and \eqref{axiom: conformal factor}, we can also follow
the following process.
\begin{enumerate}
	\renewcommand{\labelenumi}{(\arabic{enumi})'}
	\setcounter{enumi}{4}
	\item \label{axiom: geodesic}
		Solving $\alpha_{\lambda}(\theta) = \alpha_{\lambda + \epsilon v}(\theta)\,(v^0 = 0)$ to the first order of $\epsilon$ (constant),
		we obtain the tangent vector $v = v(\lambda,\theta)$ up to a normalization factor, which can be arbitrarily fixed by hand.
		The acceleration vector can be read from the coefficient of $\epsilon^1$ in the expansion of 
		$v(\lambda + \epsilon v,\theta)$.
		Putting them into the geodesic equation written in terms of the undetermined conformal factor of the metric,
		we obtain the conformal factor up to a rigid scaling, by imposing that the equation holds for any $\theta$.
	\item \label{axiom: rigid factor}
		We compute a convenient geodesic length to put into Ryu-Takayanagi formula, by which the remaining
		scaling factor is determined. 
\end{enumerate}

In the remaining of this appendix, we confirm that this alternative method works for the three examples we have dealt with in
section \ref{sec: examples}.

\subsubsection*{Pure AdS$_3$}
We restart with \eqref{eq: AdS conformal metric}.
From \eqref{eq: AdS pf}, condition $\alpha_{\lambda + \epsilon v}(\theta) = \alpha_\lambda(\theta)$ becomes
\begin{align}
	v^1\cos(\theta-\lambda^2) + v^2\lambda^1\sin(\theta-\lambda^2) = 0,
\end{align}
and hence we have
\begin{align}
	v(\lambda,\theta) = \lambda^1\sin(\theta-\lambda^2)\frac{\partial}{\partial\lambda^1}
	-\cos(\theta-\lambda^2)\frac{\partial}{\partial\lambda^1}.
\end{align}
Expanding $v(\lambda+\epsilon v,\theta)$ to $O(\epsilon^1)$, we see
\begin{align}
	a(\lambda,\theta) = \lambda^1\frac{\partial}{\partial\lambda^1} 
	+ \frac{1}{2}\sin(2(\theta-\lambda^2))\frac{\partial}{\partial\lambda^2}.
\end{align}

We put them into the geodesic equation written with $\omega(\lambda)$:
\begin{align}
	\frac{\partial\omega}{\partial\lambda^2}\tan(\theta-\lambda^2) 
	- \lambda^1\l[4\lambda^1 + \frac{\partial\omega}{\partial\lambda^1}((\lambda^1)^2-1)\r] = 0.
\end{align}
Since this holds for any $\theta$, we get
\begin{align}
	\frac{\partial\omega}{\partial\lambda^2} = 
	4\lambda^1 + \frac{\partial\omega}{\partial\lambda^1}((\lambda^1)^2-1) = 0,
\end{align}
which is solved to give
\begin{align}
	\omega(\lambda) = -\ln(1-(\lambda^1)^2)^2 + \omega_0\qquad (\omega_0\mbox{: constant}).
	\label{eq: AdS omega}
\end{align}
Therefore, we have the metric up to a rigid scaling:
\begin{align}
	\d s^2 =  e^{\omega_0}\l[
	-\frac{(\d \lambda^0)^2}{1-(\lambda^1)^2}
	+ \frac{(L\,\d\lambda^1)^2 }{\{1-(\lambda^1)^2\}^2}
	+ \frac{(L\,\lambda^1\d\lambda^2)^2}{1-(\lambda^1)^2}\r].
\end{align}

Moving to the static coordinate as we have done in subsection \ref{subsec: AdS},
we have \eqref{eq: AdS final} with it multiplied by $e^{\omega_0}$.
We know that $\omega_0 = 0$ does reproduce the proper entanglement entropy through Ryu-Takayanagi formula,
thus we omit procedure \eqref{axiom: rigid factor}' and conclude $\omega_0 = 0$.

\subsubsection*{AdS$_3$ soliton}
In this case, $v$ and $a$ are computed as follows:
\begin{align}
	v(\lambda,\theta) &= \frac{r_0}{L}\lambda^1\sin\l(\frac{r_0}{L}(\theta-\lambda^2)\r)\frac{\partial}{\partial\lambda^1}
	-\cos\l(\frac{r_0}{L}(\theta-\lambda^2)\r)\frac{\partial}{\partial\lambda^2},\\
	 a(\lambda,\theta) &= \l(\frac{r_0}{L}\r)^2\lambda^1\frac{\partial}{\partial\lambda^1} + 
	 \frac{r_0}{2L}\sin\l(\frac{2r_0}{L}(\theta-\lambda^2)\r)\frac{\partial}{\partial\lambda^2}.
\end{align}
Plugging them into the geodesic equation, we get the equation of $\omega(\lambda)$:
\begin{align}
	\frac{\partial\omega}{\partial\lambda^2}\tan(\theta-\lambda^2) 
	- \frac{r_0}{L} \lambda^1\l[4\lambda^1 + \frac{\partial\omega}{\partial\lambda^1}((\lambda^1)^2-1)\r] = 0.
\end{align}
Since this holds for any $\theta$, we again conclude \eqref{eq: AdS omega}.
The determination of $\omega_0$ is conducted by following procedure \eqref{axiom: rigid factor}', and
we conclude $\omega_0 = 2\ln(r_0/L)$ by the same reason as pure AdS$_3$.

\subsubsection*{BTZ black hole}
In this case, $v$ and $a$ are computed as follows:
\begin{align}
	v(\lambda,\theta) &= \frac{r_0}{L}\lambda^1\sinh\l(\frac{r_0}{L}(\theta-\lambda^2)\r)\frac{\partial}{\partial\lambda^1}
	+\cosh\l(\frac{r_0}{L}(\theta-\lambda^2)\r)\frac{\partial}{\partial\lambda^2},\\
	 a(\lambda,\theta) &= -\l(\frac{r_0}{L}\r)^2\lambda^1\frac{\partial}{\partial\lambda^1} - 
	 \frac{r_0}{2L}\sinh\l(\frac{2r_0}{L}(\theta-\lambda^2)\r)\frac{\partial}{\partial\lambda^2}.	
\end{align}
Plugging them into the geodesic equation, we get the equation of $\omega(\lambda)$:
\begin{align}
	\frac{\partial\omega}{\partial\lambda^2}\tanh(\theta-\lambda^2) 
	- \frac{r_0}{L} \lambda^1\l[4\lambda^1 + \frac{\partial\omega}{\partial\lambda^1}((\lambda^1)^2-1)\r] = 0.
\end{align}
Since this holds for any $\theta$, we again conclude \eqref{eq: AdS omega} (in this case we should note $\lambda^1>1$).
The determination of $\omega_0$ is conducted by following procedure \eqref{axiom: rigid factor}', and
we conclude $\omega_0 = 2\ln(r_0/L)$ by the same reason as pure AdS$_3$.

\bibliographystyle{jhep} 
\bibliography{ref}
\end{document}